\documentclass[10pt]{article}                       

\title{Analytical solution to the Schr\"odinger equation 
of a laser-driven correlated two-particle system}

\author{Uwe Schwengelbeck\\
{\em Departamento de F\'\i sica Aplicada, Universidad de Salamanca,}\\ 
{\em E-37008 Salamanca, Spain}}

\begin{document}
\maketitle

\begin{abstract}
The time-dependent quantum system of two laser-driven electrons in 
a harmonic oscillator potential,is analysed, taking into account 
the repulsive Coulomb interaction between both particles. 
The Schr\"odinger equation of the two-particle system
is shown to be analytically
soluble in case of arbitrary laser frequencies and individual 
oscillator frequencies, defining the system. 
Quantum information processing could be a possible
field of application.\\[\medskipamount]
\end{abstract}

There is an inherent interest in analytical and non-perturbative
solutions of time-dependent quantum problems, all the more 
considering quantum systems with more than one particle.
One of the few examples of such kind is the system of two
electrons in an electromagnetic field \cite{bergou}, 
where an analytical solution of the Schr\"odinger equation has 
been given recently \cite{faisal}. 
The aim of this work is to demonstrate that 
exact wavefunctions are available for the
time-dependent problem of two correlated charged particles in a 
harmonic oscillator potential, subject to a laser field
(see also Ref.\ \cite{schwe}).
The given analytical solutions \cite{schwe} are valid for
arbitrary driving laser frequencies and individual oscillator 
frequencies, defining the system. 
A possible field of application might be quantum information
processing (cf., e.g., Ref.\ \cite{roso}).
Another application of the time-dependent 
solution to the two-particle Schr\"odinger equation
could be the test of costly numerical 
algorithms to simulate the dynamics of helium
in intense laser fields---still one of the topical problems 
in theoretical atomic physics these days.
The latter system has, apart from the Coulomb interaction between 
the electrons and the nucleus, a similar structure as the 
system considered here, containing a harmonic potential. 

The wavefunction $\Psi$ of two laser-driven particles of equal mass 
$\mu$ and charge $q$ in a harmonic potential, associated with the 
oscillator frequency $\Omega$, obeys the Schr\"odinger equation
\begin{equation}\label{schroedinger}
i\hbar\frac{\partial}{\partial t} 
\Psi({\mathbf r}_1,{\mathbf r}_2, t) = 
\hat H({\mathbf r}_1,{\mathbf r}_2,t) 
\, \Psi({\mathbf r}_1,{\mathbf r}_2, t),
\end{equation}
where the Hamiltonian in length gauge \cite{delone} reads 
\begin{equation}\label{hamiltonian}
\hat H({\mathbf r}_1,{\mathbf r}_2,t) 
= \sum_{i=1}^2 \left[ \frac{\hat{\mathbf p}^2_i}{2\mu}  
+ \frac{\mu}{2}\Omega^2 {{\mathbf r}_i}^2 
- q \, {\mathbf r}_i \cdot \hat{\mathbf e} \, 
E_0 \sin(\omega t + \delta) \right] + V({\mathbf r}_1, {\mathbf r}_2),
\end{equation} 
$\omega$ and $E_0$ are the frequency and the amplitude of a
monochromatic laser field, respectively,
$\hat {\mathbf e}$ denotes a unit polarization vector, and
\begin{equation}
V({\mathbf r}_1, {\mathbf r}_2) = 
\frac{q^2}{|{\mathbf r}_1 - {\mathbf r}_2|}
\end{equation} 
describes the represents Coulomb interaction between both 
particles.

On using center of mass and relative coordinates,
\begin{equation}
{\mathbf R} = \frac{1}{2}({\mathbf r}_1 + {\mathbf r}_2)
\quad \textrm{and} \quad
{\mathbf r} = {\mathbf r}_1 - {\mathbf r}_2,
\end{equation}
the Hamiltonian (\ref{hamiltonian}) decouples,
\begin{equation}
\hat H = \frac{\hat{\mathbf P}^2}{4\mu} + \frac{\hat{\mathbf p}^2}
{\mu} +  \mu\Omega^2 {\mathbf R}^2 +  \frac{\mu}{4} \Omega^2 
{\mathbf r}^2 - 2 q\, {\mathbf R} \cdot \hat{\mathbf e} \, 
E_0 \sin(\omega t + \delta) + \frac{q^2}{|{\mathbf r}|},
\end{equation}
where 
\begin{equation}
\hat{\mathbf P} = \hat{\mathbf p}_1 + \hat{\mathbf p}_2 =
-i\hbar\nabla_{\mathbf R}, \quad \textrm{and} \quad 
\hat{\mathbf p} = \frac{1}{2}(\hat{\mathbf p}_1 - \hat{\mathbf p}_2) = 
-i\hbar\nabla_{\mathbf r}, 
\end{equation}
and the Schr\"odinger equation (\ref{schroedinger}) admits solutions 
of the factorized form
\begin{equation}\label{Psi}
\Psi = \psi({\mathbf R},t) \, \phi({\mathbf r},t) \, \chi(1,2),
\end{equation}  
where $\chi(1,2)$ is the corresponding 
spin state of the two-particle system.
The respective differential equations for the 
wavefunctions $\psi({\mathbf R},t)$ and
$\phi({\mathbf r},t) = \phi({\mathbf r}) 
\exp(-i\epsilon t / \hbar)$ in (\ref{Psi}) then read
\begin{equation}\label{CM}
i\hbar\frac{\partial}{\partial t} \psi({\mathbf R},t)
= \left[ \, \frac{\hat{\mathbf P}^2}{4\mu}  
+ \mu\Omega^2 {\mathbf R}^2 
- 2 q\, {\mathbf R} \cdot \hat{\mathbf e} \, 
E_0 \sin(\omega t + \delta) \right] 
\psi({\mathbf R},t)
\end{equation}
and 
\begin{equation}\label{rel}
\epsilon \, \phi({\mathbf r}) = \left(\, \frac{\hat{\mathbf p}^2}{\mu}  
+ \frac{\mu}{4} \Omega^2 {\mathbf r}^2 
+ \frac{q^2}{|{\mathbf r}|} \right) \phi({\mathbf r}).
\end{equation}
For both of the above equations, (\ref{CM}) and (\ref{rel}),
analytical solutions are available \cite{schwe}.

Equation (\ref{CM}), associated with the center 
of mass motion of the two particles, has essentially
the form of the Schr\"odinger equation of a driven 
harmonic oscillator.
Given an initial wavefunction $\psi({\mathbf R},0)$,
the solution of equation (\ref{CM}) can be obtained 
by means of a path integral \cite{feynman}:
\begin{equation}\label{path}
\psi({\mathbf R},t) = \int \limits_{-\infty}^{\infty} 
K({\mathbf R},t;{\mathbf R}',0) \psi({\mathbf R}',0) 
d{\mathbf R}',
\end{equation}
where the propagator, corresponding to equation (\ref{CM}), reads 
\begin{equation} 
K({\mathbf R},t;{\mathbf R}',0) =  
\sqrt{\frac{\mu\Omega}{i\pi\hbar\sin(\Omega t)}} \, 
e^{\frac{i}{\hbar} S({\mathbf R},t;{\mathbf R}',0)},
\end{equation}
containing the action function
\begin{eqnarray}
S &=& \frac{\mu\Omega}{\sin(\Omega t)} 
\left[ \, \, ({\mathbf R}'^2 + {\mathbf R}^2) \cos\Omega t \, - \,
2 {\mathbf R}' \cdot {\mathbf R} \right \delimiter 0 \nonumber\\
& & \quad + \frac{2q E_0}{\mu\Omega}\int_{0}^{t} \!
{\mathbf R} \cdot \hat{\mathbf e} \, \sin(\omega \tau + \delta)  
\sin\Omega\tau d\tau \nonumber\\
& & \quad + \frac{2q E_0}{\mu\Omega}\int_{0}^{t} \! 
{\mathbf R}' \cdot \hat{\mathbf e} \, \sin(\omega \tau + \delta) 
\sin[\Omega (t - \tau)] \, d\tau \nonumber\\
&-& \frac{2q^2 E_0^2}{\mu^2\Omega^2} \left \delimiter 0
\int_{0}^{t} \! \int_{0}^{\tau} \! 
 \sin(\omega \tau + \delta) \sin(\omega s + \delta) 
\sin[\Omega (t - \tau)] \, \sin\Omega s \, ds d\tau \right]. 
\end{eqnarray}
As an example we shall consider here a laser field with linear 
polarization along the $z$-direction and phase $\delta = 0$.
On starting with the unperturbed oscillator ground state   
\begin{equation}\label{initial}
\psi({\mathbf R},0) = \left( \frac{2 \mu\Omega}{\pi\hbar}\right)^{3/4} 
\, e^{-\frac{\mu}{\hbar} \Omega {\mathbf R}^2 }, 
\end{equation}
the center of mass wavefunction at time $t$, 
given by integral (\ref{path}), reads (${\mathbf R} = (X,Y,Z)$)
\begin{eqnarray}\label{psi}
\psi({\mathbf R},t) &=& \left ({\frac {2 \mu\Omega}{\pi\hbar}}
\right )^{3/4}
e^{-\frac{3}{2} i \Omega t} \,
e^{-\frac{\mu}{\hbar}\Omega\left ({X}^{2}+{Y}^{2}\right )} \,
e^{ i\frac{\mu}{\hbar}\Omega Z ^2 \cot\Omega t} \,
\nonumber\\
& & \quad \times
\exp\left( -\frac{\mu\Omega \left( q E_0 \, {\frac {
\sin\omega t \, - \, \frac{\omega}{\Omega} \sin\Omega t }
{\mu(\Omega^2 - \omega^2) }} - Z\right)^2}
{\hbar\sin^{2}\Omega t \,\, \left (1 - i \cot\Omega t \right )}
\right) \nonumber\\
& & \quad \times
\exp\left(2i q E_0 \, Z \,\frac{
\omega\cos\omega t \, - \, \Omega\sin\omega t \, \cot\Omega t}
{\hbar(\Omega^2 - \omega^2)} \right) \nonumber\\
& & \quad \times
\exp\left( -i q^2 E_0^2 \, \frac{\sin\omega t \, \cos\omega t
\left( \omega + \frac{\Omega^2}{\omega}
\right) - 2\,\Omega\cot\Omega t \,
\sin^{2}\omega t} {2 \mu\hbar\left(\Omega^2 - \omega^2 \right)^2} 
\right)
\nonumber\\
& & \times
\exp\left( \frac{i q^2 E_0^2 \, t}
{2\mu\hbar\left(\Omega^2 - \omega^2\right)} \right),
\end{eqnarray} 
which can be verified directly by substitution into equation 
(\ref{CM}).
It may be noted that for the case of a circularly polarized laser 
field, the wavefunction can be obtained in a similar way as in 
the above case with a linearly polarized laser field.

The remaning problem (\ref{rel}), concerning 
the relative motion of both electrons, is analytically soluble for 
an infinite denumerable set of oscillator frequencies $\Omega$ 
\cite{kais,taut1,taut2}. 
On using spherical polar coordinates, the relative 
coordinate wavefunction reads
\begin{equation} \label{polar}
\phi({\mathbf r}) = \frac{u(r)}{r}
Y_{l m}(\theta,\phi),
\end{equation} 
where $Y_{l m}$ are the spherical harmonics, and the radial part,
given by the ansatz
\begin{equation}\label{radial}
u(r) = r^{l+1} \,
e^{-\frac{\mu}{4\hbar}\Omega r^2} \, 
\sum_{\nu=0}^{\infty} a_{\nu} r^{\nu}
\end{equation}
is determined by 
\begin{equation} \label{radialeq}
\epsilon \, u(r) = \left[ -\frac{\hbar^2}{\mu}\frac{d^2}{dr^2} 
+ \frac{\mu}{4} \Omega^2 r^2
+ \frac{q^2}{r} + \frac{\hbar^2 l(l+1)}{\mu r^2}
\right] u(r).
\end{equation}
In closed form solutions of the radial equation 
(\ref{radialeq}), only a finite number 
of coefficients $a_{\nu}\neq 0$ contributes to the power series 
expansion in equation (\ref{radial}), which is satisfied in
particular cases of harmonic oscillator frequencies $\Omega$.  
As an example, an account of closed form solutions shall be given 
here for oscillator frequencies, related to the
angular momentum quantum numbers $l=0,1,\dots,\infty$ by 
(cf.\ Ref.\ \cite{taut1})
\begin{equation}\label{nu}
\Omega = \frac{q^4 \mu}{2\hbar^3(l+1)}.
\end{equation}
The eigenfunction (\ref{polar}) for the relative motion 
of the two particles is then given by
\begin{equation}\label{phi}
\phi({\mathbf r}) = r^l \, e^{-\frac{\mu}{4\hbar} \Omega r^2} 
\left(1 + \frac{\hbar}{q^2} \Omega r \right) Y_{l m}(\theta, \phi),
\end{equation} 
corresponding to an energy 
$\epsilon =  \frac{\hbar}{2}(3 \Omega + \frac{q^4 \mu}{\hbar^3})$. 
For a numerical treatment of equation 
(\ref{rel}) yielding solutions valid for arbitrary
values of oscillator frequency $\Omega$,
the reader may be refered to Refs.\ \cite{laufer,merkt}.

It is noted that the center of mass wavefunction (\ref{psi}), 
comprising the dynamical evolution of the system subject 
to the laser field, remains symmetric against particle exchange. 
At the same time, the total wavefunction 
$\Psi = \psi({\mathbf R},t) \, \phi({\mathbf r},t) \, \chi(1,2)$
has, according to the Pauli principle, 
to be antisymmetric against particle exchange. 
Thus, in case of antisymmetric solutions of equation (\ref{rel}) with
$\phi(\mathbf r) = -\phi(-\mathbf r)$, the spin state $\chi(1,2)$
of the system has to be symmetric against particle exchange,
leading to triplet states.
On the other hand, symmetric functions $\phi(\mathbf x)$ must be
accompanied by an antisymmetric spin state $\chi(1,2) = -\chi(2,1)$,
describing singlet states of the system. 
The laser interaction does not change the parity of the
system wavefunction. 
In case of a vanishing harmonic oscillator potential, i.e.\ 
$\Omega = 0$ in Hamiltonian (\ref{hamiltonian}),
the system reduces to that of two ``free'' particles in a laser field,
discussed in Ref.\ \cite{faisal}.
The relative motion of both particles is then represented by  
repulsive Coulomb wavefunctions, satisfying either
ingoing or outgoing boundary conditions. 
These solutions do not permit a motion of both particles 
with equal kinetic energy and wave vectors in the same direction
\cite{faisal}. 
In contrast to that case, the time-dependent 
center of mass wavefunction (\ref{CM}) and the stationary relative 
coordinate wavefunction (\ref{rel}) of the system, based on 
the repulsive Coulomb {\em and\/} the
harmonic potential term, describe two laser-driven particles of 
which both contributions are always pointed towards the same 
direction.

In conclusion, the time-dependent problem of two laser-driven 
correlated particles in a harmonic potential has been shown 
to be analytically soluble. 
The Hamiltonian of the Schr\"odinger equation of the 
system can be decoupled into that of a driven harmonic oscillator 
for the center of mass motion and a Hamiltonian for the 
relative motion, comprising
the repulsive Coulomb interaction between both particles. 
For both the center of mass part and the relative 
part of the two-particle problem, analytical
solutions are available.
Besides applications concerning laser interaction with ions in a trap,
and nuclear systems, the time-dependent wavefunctions 
may be utilized to check numerical algorithms to determine
the dynamical evolution of the wavefunction of two-electron 
systems, e.g.\ to simulate the dynamics of the helium atom
in laser fields. 
Another possible field of application might be quantum information
processing.

\section*{Acknowledgments}
It gives me great pleasure to thank Luis Plaja, Luis Roso,
Javier Rodr\'\i guez V\'azquez de Aldana
and Roberto Numico for the stimulating exchange
and useful discussions.

\end{document}